\documentclass[smallextended]{svjour3}

\let\phi=\varphi

\usepackage{amsfonts}
\usepackage{fancybox}
\usepackage[hyphens]{url}
\usepackage[utf8]{inputenc}
\usepackage{xspace}
\usepackage{dcolumn}
\usepackage{rccol}
\usepackage{ifpdf}
\usepackage[pdftex,bookmarks=true]{hyperref}

\def\systemname#1{\textsf{#1}\xspace}
\def\libname#1{\textsf{#1}\xspace}

\newcommand{\mizar}{\systemname{Mizar}}

\newcommand{\mml}{\libname{MML}}

\newcommand{\coq}{\systemname{Coq}}

\newcommand{\TeXmacs}{T\kern-.1667em\lower.5ex\hbox{E}\kern-.125emX\kern-.1em\lower.5ex\hbox{\textsc{m\kern-.05ema\kern-.125emc\kern-.05ems}}\xspace}

\usepackage{xcolor}

\usepackage{listings}
\newcommand{\term}[1]{\textbf{#1}}

\def\ourtitle{Eliciting implicit assumptions of proofs in the \mizar{} Mathematical Library by property omission}

\hypersetup{
  pdftitle={\ourtitle},
  pdfauthor={Jesse Alama},
  pdfkeywords={interactive theorem proving, proof dependencies, proof analysis, large formal libraries, mizar},
  colorlinks=true,
  linkcolor=black,
  citecolor=black,
  breaklinks=true,
}

\title{\ourtitle \author{Jesse Alama$^{1}$\thanks{The author was supported by the FCT project
    ``Dialogical Foundations of Semantics'' (DiFoS) in the ESF
    EuroCoRes programme LogICCC (FCT LogICCC/0001/2007).}}}
\institute{Center for Artificial Intelligence\\
  New University of Lisbon\\
  \email{j.alama@fct.unl.pt}
}
\authorrunning{Alama}
\titlerunning{Eliciting implicit assumptions in \mizar{}}

\begin{document}

\maketitle

\begin{abstract}
  When formalizing proofs with interactive theorem provers, it often
  happens that extra background knowledge (declarative or procedural)
  about mathematical concepts is employed without the formalizer
  explicitly invoking it, to help the formalizer focus on the relevant
  details of the proof.  In the contexts of producing and studying a
  formalized mathematical argument, such mechanisms are clearly
  valuable.  But we may not always wish to suppress background
  knowledge.  For certain purposes, it is important to know, as far as
  possible, precisely what background knowledge was implicitly
  employed in a formal proof.  In this note we describe an experiment
  conducted on the \mizar{} Mathematical Library of formal
  mathematical proofs to elicit one such class of implicitly employed
  background knowledge: properties of functions and relations (e.g.,
  commutativity, asymmetry, etc.).
\end{abstract}

\section{Introduction}

When formalizing mathematical proofs with interactive theorem provers,
it often happens that extra background knowledge is employed without
the formalizer explicitly invoking it.  The effect is clear: thanks to
such facilities, the formalizer can focus more on the relevant details
of the proof he is working on, rather than (relatively more) tedious
``details''.  In the contexts of producing and studying a formalized
mathematical argument, such mechanisms are important and deserve to be
strengthened.

But we may not always wish to suppress background knowledge.  For
certain purposes, it is important to know, as far as possible,
precisely what background knowledge was implicitly employed in a
formal proof.  There are a number of practical applications of such
information, such as facilitating theory exploration~\cite{proofs-and-refutations} , library
recompilation~\cite{large-formal-wikis}, improved machine learning about formal proofs, premise selection
heuristics, etc.  The aim of eliciting exactly what is used is also
important philosophically, because we come closer to finding what is
truly necessary for the success of a theorem.  Our project thus aligns
in a modest way with the overall aims of reverse
mathematics~\cite{simpson-sosoa}, a vibrant branch of contemporary
proof theory whose aim is to discover axioms from theorems (rather
than the other way around).

The task of eliciting all implicit information of a formalized
mathematical proof is a stimulating challenge.  Various approaches can
be used depending on the ITP and its associated logic(s).  (For a
comparison of how this can be done for \mizar{} as compared to \coq,
see~\cite{alama-mamane-urban}.) In this note we describe an experiment
conducted on the \mizar{} Mathematical Library (\mml) of formal
mathematical proofs to elicit one such class of implicitly employed
background knowledge: properties of functions and relations (e.g.,
commutativity, asymmetry, etc.).  We employ the term
\term{constructor} to mean function or relation in a formal
mathematical library.\footnote{The term ``constructor'' is in fact part of the \mizar{} idiolect for formal mathematics.  There are more kinds of objects that count as ``constructors'' in \mizar{}---such as structures---but in this paper we are interested only in functions and predicates.}  The \mml{} is a large collection of formal
mathematical proofs expressed in classical first-order set theory and
a natural deduction-style declarative proof formalism.  (For an
introduction to \mizar, see~\cite{mizar-in-a-nutshell}.)  Analogous
experiments could clearly be performed on other libraries of
formalized mathematical knowledge.  The \mizar{} system is especially
attractive for this kind of experiment because of the clear semantics
of the properties that can be implicitly attached to functions and
relations (they are simple universal first-order formulas) and the
relative ease of manipulating these properties.

Section~\ref{sec:previous-work} outlines some previous related work
and sketches some of the intended applications of the constructor
property dependency data.  Section~\ref{sec:omitting} gives two
examples of how constructor properties can be implicitly used in
formal proofs in \mizar{}.  Section~\ref{sec:eliciting} describes the
method we used to make explicit the constructor properties that are
needed for \mizar{} inferences. The heart of the paper is
Section~\ref{sec:usage}, where we give the results of our
computation about needed and unneeded constructor properties for the
entire \mizar{} Mathematical Library.

\section{Applications and previous work}
\label{sec:previous-work}

\mizar's notion of constructor properties (and another mechanism for
suppressing the premises of an inference, so-called requirements) are a relatively new invention for \mizar~\cite{DBLP:conf/mkm/NaumowiczB04}.  The information
that a constructor property is needed for an inference can be
exploited in various ways.  One potential application is to use the
needed-property information as an indication of what formula shapes
are useful in the search for a successful proof.  Building on the
ideas of S.~Shulz~\cite{Schulz:Diss-2000}, one would strip away the
actual symbols employed in a proof that implicitly relies on some
constructor property, thereby learning some structural information
about what contributes to the success of a theorem.  One could
profitably using this information to assist in the problem of
selecting candidate premises in large ATP
problems~\cite{premise-selection}.  More generally, such a structural
approach can give us some information about shapes of successful
patterns of formal reasoning.

The information that a property of a constructor is not needed
indicates, prima facie, that one is dealing with a kind of
generalization.  It may not always be clear, though, how to use this
information to craft generalizations. The problem problem lies in the
familiar distinction in logic between reasoning formally about a class
of structures or about a single structure.  It would seem that
information that a property is not needed information is more useful
in cases of reasoning about classes of structure (e.g., about fields),
but in the case of reasoning about ``concrete'' mathematical objects
(a single structure, e.g., the real numbers).  Thus, we may find in a
theorem about real numbers that commutativity of addition is not used.
One could immediately construct a generalization of such a theorem to
a class of structures that are just like the real numbers, but in
which the commutativity of addition is not assumed.  The value of such
a generalization may not be clear.  By contrast, imagine we encounter
a theorem about fields in which the commutativity of the field's
addition operation is not used.  In this case, the generalization
procedure is similar to the procedure used in the case of the reals.
By contrast, though, the significance and utility of such a
generalization is clearer because the reasoning was already about a
wide class of structures, the properties of whose functions and
relations enjoyed some flexibility.

One might imagine an advisor attached to an interactive theorem prover
that can use the information that a constructor property is needed to
help one investigate and formulate suitable generalizations.  As discussed earlier,
one relevant task (which seems quite interesting from the perspective
of artificial intelligence) would be to decide whether a
generalization is even warranted.  It's intuitively clear that when we
use a mathematical concept in a proof we use only some aspects of it
and not others; at some steps, we require one thing of a concept, but
at other steps we require something else.  Dividing a concept so that
we never require from it nothing less than its ``full content'' every
time it is used seems like a drastic suggestion.  Still, we may find
value in a mitigated form of this advice.


\section{Omitting properties of functions and relations}
\label{sec:omitting}

In \mizar{}, one can \term{attach} properties to functions and relations;
Tables~\ref{tab:properties-of-relations} and~\ref{tab:properties-of-functions} lays out all nine properties
currently supported by \mizar.  These properties are used by the
\mizar{} verifier as ``background knowledge'' that doesn't need to be
cited by a formalizer.
\begin{table}\normalsize
  \centering
  \begin{tabular}{cc}
Relation Property & Formula\\
\hline
reflexivity       & $\forall x [R(x,x)]$\\
symmetry          & $\forall x \forall y [R(x,y) \rightarrow R(y,x)]$\\
asymmetry         & $\forall x \forall y [R(x,y) \rightarrow \neg R(y,x)]$\\
connectedness     & $\forall x \forall y [R(x,y) \vee R(y,x)]$\\
irreflexivity     & $\forall x [\neg R(x,x)]$\\
\end{tabular}
\caption{Properties of relations in \mizar}
\label{tab:properties-of-relations}
\end{table}

\begin{table}\normalsize
  \centering
  \begin{tabular}{cc}
Function Property & Formula\\
\hline
projectivity      & $\forall x [f(f(x)) = f(x)]$\\
involutiveness    & $\forall x [f(f(x)) = x]$\\
idempotence       & $\forall x [g(x,x) = x]$\\
commutativity     & $\forall x \forall y [g(x,y) = g(y,x)]$
\end{tabular}
\caption{Properties of functions in \mizar}
\label{tab:properties-of-functions}
\end{table}

\subsection{Example 1: Relation property}
\label{sec:relation-property}

Consider the definition of the proper subset relation:
\begin{lstlisting}
definition
 let X,Y be set;
 pred
  X c< Y
 means
  X c= Y & X <> Y;
 irreflexivity;
 asymmetry;
\end{lstlisting}
This example, taken from the \mizar{} article \texttt{XBOOLE\_0},
defines the predicate (\texttt{pred}) of one set \texttt{X} being a
proper subset \texttt{c<} of another \texttt{Y}.  The symbols
\texttt{c=} and \texttt{<>} denote the subset relation and
disequality, respectively.  The keywords \texttt{irreflexivity} and
\texttt{asymmetry} included in the definition indicate that the proper
subset relation will henceforth have the properties of irreflexivity
and asymmetry; inferences involving the proper subset relation will
implicitly use these properties.

There are 58 occurrences in the \mizar{} Mathematical Library in which
the irreflexivity of the proper subset relation is implicitly used.
Here is one example, taken from the article \texttt{TREES\_1}:
\begin{lstlisting}
theorem
not <*n*> is_a_proper_prefix_of <*m*>
proof
assume A1: not thesis;
then <*n*> is_a_prefix_of <*m*> by XBOOLE_0:def 8;
hence contradiction by A1,Th16;
end;
\end{lstlisting}
The theorem says that the one-term sequence whose sole term is the
number $n$ is not a proper prefix of the one-term sequence whose sole
term is the number $m$.  The binary relation
\texttt{is\_a\_prefix\_of} on finite sequences is defined simply as
set inclusion.  The reference \texttt{Th16} refers to the theorem that
says that a one-term finite sequence $\langle s \rangle$ is a prefix
of the one-term finite sequence $\langle t \rangle$ iff $s = t$.  The
proof of our theorem goes by contradiction.  Note that the symbol for
the proper subset relation, \texttt{c<}, occurs neither in the theorem
nor in the proof.  In the absence of the assumption that \texttt{c<}
is irreflexive, the contradiction at the end of the proof does not
follow.  A contradiction indeed does not follow: we get from
\texttt{Th16} that $n = m$, but this is compatible with the one-term
sequence \texttt{<*n*>} being a proper subset of \texttt{<*m*>}, if we
haven't assumed that the proper subset relation is irreflexive.

\subsection{Example 2: Function property}
\label{sec:function-property}

Additive magmas come equipped, of course, with an addition operation
$+$.  In the case of abelian additive magmas, we know that $+$ is
commutative:
\begin{lstlisting}
definition
 let V be Abelian addMagma,
     v be Element of V,
     w be Element of V;
 redefine func v + w;
 commutativity;
end;
\end{lstlisting}
This example is taken from the \mizar{} article \texttt{RLVECT\_1}.
The keyword \texttt{redefine} here does not indicate that we are
changing the definiens of the binary function $+$ on elements of
additive magmas (which in any event is essentially undefined); rather,
we are attaching the property of commutativity to $+$.  Such an
operation is obviously admissible because of the definition of what it
means for an additive magma to be abelian.

As an example of an inference that implicitly depends on the
commutativity of $+$, consider:

\begin{lstlisting}
theorem
 for L being add-associative right_zeroed
             right_complementable Abelian
             non empty addLoopStr,
  b, c being Element of L
  holds c = b - (b - c)
proof
  let L be add-associative right_zeroed
           right_complementable Abelian
           non empty addLoopStr,
      b, c be Element of L;
  set a = b - c;
  a+c-a = c-a+a by RLVECT_1:28
       .= c by Th1;
  hence thesis by Th1;
end;
\end{lstlisting}
Ignoring the definition of all the attributes that are being attached
to the type \texttt{addLoopStr} (viz., an additive Moufang loop
structure), the crucial step here is the equation on line 17.  Note
that the terms \texttt{a} and \texttt{c} are being swapped without
reference to the commutativity of \texttt{+} (the reference to the
theorem \texttt{RLVECT\_1:28} is not relevant here).

\section{Eliciting needed implicit constructor properties}
\label{sec:eliciting}

To elicit the constructor properties that are needed for an item of
the \mizar{} Mathematical Library, we exploit \mizar{}'s separation of
(i) the construction of the environment in which verification will be
carried out from (ii) the process of verification properly speaking.

In step (i), \mizar{} constructs an environment for verification,
importing all constructors that occur explicitly or implicitly in a
\mizar{} text.  If a constructor has a property associated with it,
the environment will contain the property attached to the constructor,
regardless of whether it is truly needed.  The environment is thus a
conservative overestimate of what is truly needed for the verification
to succeed.  By intervening between the construction of the
environment and the verification proper, one can manipulate the
environment in which the verification is carried out.  We simply
remove a property attached to a constructor and carry out the
verification; if the verification succeeds, we know that the property
of the construction was not actually needed.

Thanks to the use of XML as the representation of the environment for
\mizar{} articles~\cite{DBLP:conf/mkm/Urban05}, conducting our
experiment is as simple as applying certain XSL stylesheets to the
environment files.

Rather than operating on whole \mizar{} articles, which generally
contain dozens if not hundreds of toplevel items, we operate on a
individual theorems of the \mml.  This is made possible by dividing
the \mml{} into fine-grained ``items'' (which are in fact valid
\mizar{} ``microarticle''); see~\cite{alama-mizar-items} for more
information on how this decomposition of the \mml is carried out.

\section{Usage of properties throughout the MML}
\label{sec:usage}

We have so far said that a verifiable item $I$ of the \mizar{}
Mathematical Library depends on property $P$ of constructor $C$ just
in case $I$, in the absence of the attachment of $P$ to $C$, is not
verifiable.  This definition of dependence upon a constructor property
fails to capture the dependence of one item upon another.  For
example, suppose that an item $I$ does not depend on property $P$ of
constructor $C$, that is, $I$ is verifiable if one detaches from $C$
the property $P$.  Suppose, though, that item $I$ depends on some
other item $I^{\prime}$ which \emph{does} need property $P$ of $C$.
Such a though experiment suggest that we distinguish two sense of
``need'': direct and indirect.

\begin{definition}
  Item $I$ \term{directly needs} property $P$ of constructor $C$ iff
  verification of $I$ will fail if $P$ is detached from $C$.

  Item $I$ \term{indirectly needs} property $P$ of $C$ iff $I$
  directly needs $P$ of $C$ or there exists an item $I^{\prime}$ such
  that $I$ depends on $I^{\prime}$ and $I^{\prime}$ needs $P$ of $C$.
\end{definition}

Both senses of ``need'' are useful.  Table~\ref{tab:need-statistics}
gives statistics about various constructor properties that are needed
directly and indirectly in the \mml.

\begin{table}\normalsize
  \centering
  \begin{tabular}{ccc}
  Property & Direct Items & Indirect Items\\
  \hline
  reflexivity & 54113 & 102426\\
  symmetry & 29744 & 97220\\
  asymmetry & 256 & 82585\\
  connectedness & 5020 & 83083\\
  irreflexivity & 91 & 65951\\
  projectivity & 153 & 10002\\
  involutiveness & 533 & 67853\\
  idempotence & 535 & 70132\\
  commutativity & 14055 & 92580\\
\end{tabular}
\caption{Direct and indirect dependence upon properties in the \mml}\label{tab:need-statistics}
\end{table}

There are a number of fascinating details behind these statistics:

\begin{itemize}
\item Reflexivity is directly needed by nearly half of the items of
  the \mizar{} Mathematical Library, and is indirectly needed by
  nearly the entire library.  Reflexivity of equality of sets accounts
  for this: it is indirectly needed by fully 102242 items.  It is
  perhaps not surprising that such a fundamental property of a
  built-in logical notion pervades the library.

  Putting aside equality, the next most important reflexive
  constructor is subset inclusion, which is indirectly needed by 93284
  items.  A redefinition of subset inclusion for ordinals is
  indirectly used by 8279 items.  Putting aside these ``logical'' or
  ``set theoretical'' examples, the most important reflexive
  ``mathematical'' relation is the less-than-or-equal-to relation
  $\leq$ on (extended) real numbers, whose reflexivity is indirectly
  needed by 67196 items.
\item Irreflexivity is directly needed by only a handful of items in
  the library, but indirectly it supports about 2/3 of the library.
  The explanation is the proper subset relation: the irreflexivity of
  this constructor is needed by 65546 items.  The most important
  ``mathematical'' example is the relation of one element of a
  relation being strictly less than another.
\item Asymmetry is attached to only five constructors in the entire \mml:
  \begin{itemize}
  \item $\in$ (set membership);
  \item the proper subset relation;
  \item a variant of $\in$, defined for many-sorted set structures;
  \item the strictly-lexicographically-less-than relation on finite
    tuples of natural numbers;
  \item the strictly-lexicographically-less-than relation on bags of
    ordinals.
  \end{itemize}
\item The asymmetry, of $\in$, is foundationally significant in the
  sense of mathematical logic (it expresses a weak form of the axiom
  of foundation) and for the \mizar{} Mathematical Library.  The
  asymmetry of this constructor alone accounts for essentially all
  items that need the asymmetry of any constructor: 82581 items
  indirectly need this weak form of foundation, whereas only 283 items
  directly depend on this property of $\in$.\footnote{This is arguably
    a curiosity of the organization of the \mml.  It turns out that
    asymmetry of $\in$ is used very early on in the logical
    construction of the \mml, hence its outsized influence.}
\item Likewise, projectivity is rarely directly needed, but supports a
  substantial piece of the library.  Interestingly, it is the closure
  operation defined on subsets of a topological space\footnote{To be
    precise, the operation is defined not on topological
    \emph{spaces}, but on topological \emph{structures}, which one can
    think of simply as a class of structures that has a carrier and a
    topology, viz., a collection of subsets of the carrier.  A
    topological structure on its own is not assumed to have any
    properties beyond these.} that accounts for the lion's share of the
  items that indirectly need an projective constructor: 7536 items
  depend on the projectivity of the closure operation.
\item Connectedness is attached to very few constructors of the
  \mizar{} Mathematical Library, but the constructors to which this
  property is attached have a significant influence across the \mml.
  The constructor whose connectedness is used indirectly by the
  greatest number of items is the subset relation, restricted to
  ordinals.  The connectedness of this constructor expresses a rather
  significant fact about ordinals (any two ordinals are comparable).
  The proof of this fact in \mizar{} uses a trichotomy-like principle
  for ordinals, saying that for any two ordinals $A$ and $B$, either $A
  \in B$, $A = B$, or $B \in A$.  The connectedness of the subset
  relation on ordinals is indirectly needed by 82490 items.  The next
  most significant example is $\leq$ on rational numbers, indirectly
  needed by 71313 items.
\item Involutiveness also requires some explanation.  There are two
  items that vie for the most important here:
  \begin{itemize}
  \item The sign-changing operation $x \mapsto -x$ on real numbers is
    needed by 65501 items.
  \item The reciprocal operation $z \mapsto 1/z$ on complex numbers is
    needed by 65105 items.
  \end{itemize}
  The constructor with the next highest number of items that
  indirectly depend on its involutiveness is the relative complement
  operation, which is indirectly needed by 8847 items.
\item The constructor whose idempotence is most frequently needed
  indirectly is the binary union of two sets, which is indirectly
  needed by 69184 items.  The idempotence of binary set intersection
  takes second place: it is indirectly needed by 24249 items.
\item Several items in the \mizar{} Mathematical Library indirectly
  depend on more than 100 constructor properties.  The item with the
  greatest number of dependencies is taken from the proof of the
  Jordan curve theorem:
  \begin{lstlisting}
theorem
 for n being Element of NAT
 for C being being_simple_closed_curve
  Subset of (TOP-REAL 2)
   st n is_sufficiently_large_for C
   holds cell(Gauge(C,n),
              X-SpanStart(C,n) -'1,
              Y-SpanStart(C,n))
     misses C
proof
  assume n is_sufficiently_large_for C;
  then cell(Gauge(C,n),
            X-SpanStart(C,n) - '1,
            Y-SpanStart(C,n)) c= BDD C by Th6;
  hence thesis by JORDAN1A:15,XBOOLE_1:63;
end;
  \end{lstlisting}
  The precise mathematical definitions, the theorem references
  (\texttt{Th6}, \texttt{JORDAN1A:15}, etc.) and the proof here are
  not important.  What is important is that we are dealing here with a
  very short proof (it has only three steps) of a theorem along the
  way to a substantial landmark in formal mathematics.  Despite
  appearances, this theorem indirectly depends on fully 113
  constructor properties.

  The formalization in \mizar{} of the Jordan curve theorem required a
  great deal of work and made heavy use of the \mizar{} system's
  features, such as its support for constructor properties under
  discussion here.  We see this by virtue of the fact that the
  theorems of the series of \mizar{} articles leading to the final
  proof of the Jordan curve theorem indirectly need, on average,
  several dozen constructor properties.
\end{itemize}

\section{Conclusion and future work}
\label{sec:conclusion}

The abstractions discovered here in the context of an interactive
theorem prover could, in all probability, be discovered equally well
by an ATP.  With an automated theorem prover one could, in principle,
go much further than we have gone here.  An infrastructure for carrying
out such exploration (sound translation of \mizar{} proofs to a
vanilla unsorted first-order language, infrastructure for constructing
and working with the associated ATP problems, etc.) already
exists~\cite{urban2006}.  One could even verify the dependency claims
made here outside of \mizar, in the style of~\cite{Urban-GDV}.

When we discover that some property of a function or relation is
needed, we are discovering not that the property is logically or
mathematically needed for the success of the theorem in question;
instead, we are discovering only that there exists a formal deduction
(viz., a \mizar{} proof) where this property is used.  It is certainly
quite possible that there would exist other formal proofs of the same
theorem that do not use the property.  This suggests a much more
ambitious project: discovering \emph{new proofs} of theorems that use
much less than what an interactively constructed proof uses.  Such
prospects are enticing, and deserve to be carried out.  Such a project
naturally also bears on the philosophical problem of when two proofs
are ``the same'': it could very well be that there are two proofs of
the same theorem, one which exploits a property of a function or
relation, and another which doesn't, but which should nonetheless,
from another perspective, be considered identical.

Constructor properties are but one mechanism in \mizar{} that hides
premises of inferences.  \mizar{} also supports so-called
requirements, which also help to allow one to reason validly without
having to be explicit about precisely what premises are needed for
every inference of a proof.  \mizar{} also has some built-in
functionality concerning arithmetic.  Both of these mechanisms are of
great value for the formalizer when constructing a formal proof, but
if one is interested in making explicit premises that were suppressed
during proof construction, \mizar{}'s requirements and arithmetic
facilities need to be taken into account.

One might wonder why there are only the nine constructor properties
supported by \mizar{}.  Binary relation transitivity, unary function
surjectivity and injectivity, for example, are conspicuously absent.
Supporting such additional function and relation properties could be
quite valuable; \mizar{} itself and tools using its library could
exploit function injectivity and surjectivity, for example, to help
rule out the solution for certain search problems that require the
domain of discourse to be finite (see~\cite{finite-unsatisfiability}).
But more generally, it would be valuable to mine the \mizar{}
Mathematical Library for common shapes of formulas that play a large
inferential role, and which could naturally be promoted to the level
of constructor properties.  One might discover fruitful properties
that could help make formalization in \mizar{} even more appealing and
practical.

\bibliographystyle{splncs03}
\bibliography{mlpa}

\end{document}